# Analysis of Gamma Rays and Cosmic Muons with a Single Detector

A. Bachri[1], P.C. Grant[1], and A. Goldschmidt[2]

[1]Southern Arkansas University, Magnolia, AR 71754, [2]Lawrence Berkeley National Laboratory, Berkeley, CA 94720

Correspondence: agbachri@saumag.edu

## Abstract

In this paper, we report on the construction and upgrade of a 2002 Lawrence Berkeley National Laboratory (LBNL) Quanknet Cosmic Muons Detector. By adapting this model, we modify the electronics and mechanics to achieve a highly efficient gamma-ray and cosmic-ray detector. Each detector module uses a one-inch-thick scintillator, attached to a photomultiplier tube (PMT) and mounted on a solid aluminum frame. A mechanical support was designed to allow flexible positioning between the two modules. The detector uses scintillation to transform passing radiation into detectable photons that are guided toward a photocathode surface of the PMT, triggering the release of photoelectrons that are then amplified to yield measurable electronic signals. The modules were connected to an electronics section that compared the signals from the two PMTs and logically determined if they were coincidence events. A data-collection device was added for faster count rates and to enable counts for extended times ranging from a few hours to days as needed. Count rates were taken at a variety of distances from the radioactive source, $^{60}$Co (cobalt), which produced two gamma rays and a beta particle. To investigate the isotropic behavior of radiation, two detection modules were adjusted to different angles of rotation with respect to each other, and the coincidence counts were measured. The coincidence counts from the modules set at various angles were consistent throughout the angular spectrum, and only lead shielding visibly reduced the number of counts from the radioactive source. The inverse-square-law behavior of radiation has also been considered. The results were such that the number of counts decreased as a function of increasing distance from the source. Furthermore, positioning the detector to point toward the sky in different orientations, we measured cosmic-ray muon flux as the angle from the vertical was decreased. In doing so, we scanned different patches of the atmosphere. For the optimum operation during the detection phase, we plateaued both PMTs to single out their best operating gain voltage while eliminating false background noise signals. The detector is more efficient and adaptable in collecting both gamma rays and cosmic-ray muon-flux information.

## Introduction

The inhabitants of the earth are constantly exposed to radiation that comes in two forms: charged-particle radiation and uncharged radiation. It was realized early on that radiation increased rapidly with altitude, suggesting that it had extraterrestrial origins. This hypothesis was confirmed by Robert Andrews Millikan (Millikan 1947), who gave the radiation the name of Cosmic Rays (CRs). DNA mutation and damage, which are implicated in many cancers and genetic disease, can be caused by high-energy CRs (Francis et al. 2006) therefore making the study and understanding of these cosmic rays very important.

CRs are charged particles, with approximately 89% of these particles being protons that have been accelerated to near the speed of light by unknown mechanisms. The origin and nature of acceleration mechanisms of these particles is one of the oldest unanswered fundamental questions in particle physics and cosmology. The highest energy cosmic rays observed have energies in the range of $10^{20}$ eV and thus provide a glimpse of physics and astrophysics at energies unattainable in laboratory experiments (Günter 2001). Through collision processes, the protons interact with nitrogen, oxygen and other atoms in the Earth's upper atmosphere, and numerous secondary particles are produced, forming CRs showers on earth (Bhattacharjee 2000).

In these high-energy collisions, such as $p + n \rightarrow p + p + \pi^-$ and $p + n \rightarrow p + n + \pi^0$, secondary high-energy pion particles, which belong to the meson family, are produced. Pions decay rapidly, with a mean life of 26 ns for charged pions ($\pi^\pm$) and $10^{-16}$ s for uncharged pions ($\pi^0$). A high-energy (charged) pion decays via weak interactions to muons and neutrinos as $\pi^+ \rightarrow \mu^+ + \nu_\mu$ and $\pi^- \rightarrow \mu^- + \bar{\nu}_\mu$. The muon is an electrically charged particle that is similar to an electron, but has roughly 200 times its mass and a lifespan of about 2.2 µs, the second-longest lifetime of all known particles. Muons have two properties that allow them to reach the earth's surface: they decay relatively slowly compared to pions, and they penetrate large amounts of material due to their large energy. Muons, unlike pions, have no strong interaction properties and unlike electrons they are too massive to be significantly deflected by atomic electric fields they may encounter.

The other form of cosmic rays is uncharged, electromagnetic radiation that comes in various forms, from radio waves used in cell phones to gamma rays used in radiation treatments for cancer patients. In the spectrum of electromagnetic radiation, gamma rays have the most energy and the shortest wavelength. Gamma rays are produced in the universe by supernovae explosions, solar flares, neutron stars, black holes, and active galaxies. Gamma-rays coming from space are mostly absorbed by the Earth's atmosphere. The most common means of production of gamma rays for scientists to study is the beta decay of certain isotopes. Radioactive decay is the spontaneous disintegration of a radionuclide, accompanied by the emission of ionizing radiation in the form of alpha or beta particles or gamma rays, which are responsible for the majority of background radiation that all life on earth is exposed to on a daily basis.

The current paper focuses on adapting an otherwise CR detector to detect gamma rays, and improving the efficiency of CR detection. Before the detector was modified to accommodate gamma rays, the gamma rays would penetrate the scintillators with little or no reaction at all. By increasing the thickness of the paddles, the chance of slowing and detecting a gamma ray was achieved. In the following sections, we overview the construction phase of the detector, and describe its basic principle and the plateauing procedure. To ensure the overall consistency and functionality of the machine, we investigated the radiation of gamma rays from an active $^{60}$Co source[1], demonstrated how the radiation changes with increased distance from the source, and studied its isotropic behavior and the effects of different shielding on the gamma rays produced, furthermore, we investigated CR muon flux dependence on direction; we do not expect the flux to be isotropic because of atmospheric loss through decay of muons with long tracks. We hope the detector becomes standard physics-laboratory equipment so that the physics students of tomorrow will have a greater tool to investigate the nature of radiation that permeates our world of cosmic origin.

**Materials and Methods**

*i. Adaptation to Previous Model*

The LBNL "CR detector" (Collier and Wolfley 2006) was originally built in 2002. It consists of two scintillators, light guides, PMTs, and a circuit board. Unfortunately, it has been limited to only studying CR muons. The research group from Southern Arkansas University used the previous LBNL CR detector as a basis for constructing a more versatile, portable muon detector and an efficient gamma-ray detector. This part was accomplished by adopting thicker EJ-212 scintillators to improve gamma ray efficiency. Rather than cutting scintillation material to fit the window of a PMT, a light guide made of acrylic plastic (Lucite) served as optical medium between the scintillator and PMT. By total internal reflection, the efficiency of detection was improved. In reality, efficiency of light transmission from the scintillator to the PMT window via a light guide depends greatly on the surface area of the optical couplings between the three components. In the case of this modified detector, photons-transmission efficiency was no greater than 33 percent. Without the light guide, and with simply air as a transition medium, decreased efficiency of light transmission would be unavoidable.

In addition, we separated the electronics, and used voltage control for the PMT gain and an external pulse counter with a USB interface connected to a laptop computer. This external pulse counter allowed for an increased number of counts and record time; however, the hardware is fully functional, with a built-in counter (although limited to 60-second counts), but it eliminated the need of an external pulse counter for some types of investigations.

*ii. Scintillators*

A scintillator is a substance that produces light when traversed by charged particles. It relies on the property of materials, such as thallium - activated sodium iodide, which emit light when bombarded with ionizing radiation. We can use this emitted light as a trigger whenever radiation passes through the scintillators. Scintillators can be made from various materials, such as crystals, plastics, and semiconductor materials. We used an EJ-212 plastic scintillator manufactured by Eljen Technology. The scintillator is composed of a polyvinyl toluene base that has been doped with dye molecules, 2-methylnapthalene ($C_{11}H_{11}$) (Eljen Technology 2009). This light travels through the scintillator and is channeled to the PMT by total internal reflection. The scintillation efficiency of the scintillators is 10000 photons for every 1 MeV of energy deposited.

---

[1] $^{60}$Co decays to the stable element $^{60}$Ni, and as it decays it produces a beta particle and two gamma rays with a total energy of 2.5 MeV.

### iii. Photomultiplier Tubes

The PMTs were connected to the scintillator via an acrylic light guide that was shaped, polished and glued to both the scintillator and PMT glass surface, thus forming a paddle-shaped construction. This unit works on the principle of the famous photoelectric effect first proposed by Einstein.

A PMT consists of an input window, a photocathode, focusing electrodes, an electron multiplier, and an anode usually sealed inside an evacuated glass tube. When a photon of light hits the photocathode of low work function, the photocathode produces an electron, therefore converting photonic signals into measurable electronic signals. The electron-multiplier section consists of dynodes, which amplify the electrons by way of secondary electron emission until there is a sufficient number of electrons to produce a several-hundred-millivolt signal. The detector that we constructed uses a P30CW5 PMT (Hamamatsu 2006) that incorporates a blue-green sensitive bi-alkali photocathode and a multi-stage high-voltage (HV) power supply. This voltage is applied to a chain of dynodes and ranges from 300 V to 1800 V, more than 1000 times the power delivered by the PMT's power distribution unit. Only photons emitted within the scintillators are desirable for accurate measurements; therefore, we carefully wrapped the scintillators, light guide, and PMT together with light-tight materials that block unwanted visible light (noise) from entering and flooding the sensitive PMT surface.

### iv. Electronics Section

The electronics section of the detector processes the pulses from the PMTs and converts them into counts that are displayed on a readout. Another smaller circuit board was designed by the team and incorporated into the electronics section of the detector to adjust the control/gain voltage to the PMTs. The plateauing of the PMTs, which indicates the optimal operating voltage, allows for greater sensitivity with little interference of background radiation. The collection of resistors, capacitors, diodes and integrated circuits compares PMT output pulses against a given voltage. This voltage difference can be as small as 7 mV and still make a detection event. Then the electronic circuit determines, through a series of logic gates, if the pulses from each PMT are close enough in time to call the pulses coincidences. The electronics require that hits in each module occur within $8 \times 10^{-7}$ s to qualify as a count; only then will the electronic circuits advance the counter. The circuit board also features an RC timer that can be set to a given time, chosen to be 1 minute, so that repeated measurements can be taken. This may be a buzzer that is activated each time a hit is registered.

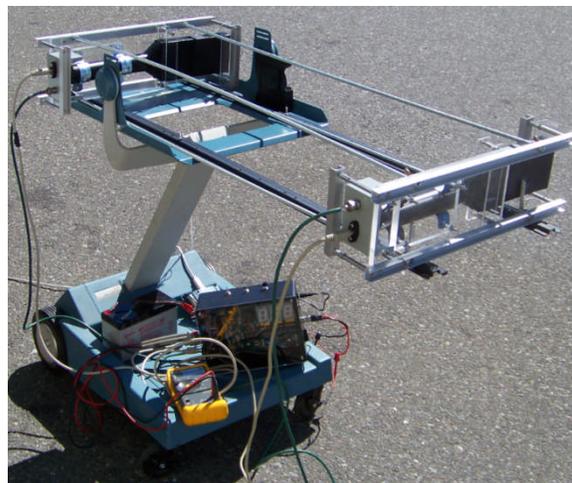

Figure 1. The cosmic muon and gamma-ray detector. The configuration shown is for measuring muon flux as the detector is pointed to the sky at a given angle (90° with respect to the vertical here).

### v. External Pulse Counter

The external pulse counter, although not required, allows the CR detector greater flexibility. We used a USB-1208LS personal measurement device with digital and analog inputs and outputs in addition to a counter function. This commercially available plug-and-play device, compatible with TracerDAQ software, allowed for data from the CR detector to be fed directly to a computer, and enabled event counting that may require long time measurements or that may run too fast to count via the onboard counter. During the course of our data collection the USB interface permitted overnight recording of events when required. The computer program supplied with the USB pulse counter was used to record all the data from the pulse counter.

### vi. Methods

Once the detector was fully functional and the research team had tested it, a series of calibration procedures was performed. These procedures consisted of both single PMT pulse rates and coincidence rates to calibrate the detectors for muons and gamma rays. Each sample of single counts was taken in one minute intervals and ranged from 0 counts at the lowest setting of control voltage to over 60000 counts at the highest setting. The large variation of counts with respect to the PMT operating voltage produced plateau regions.

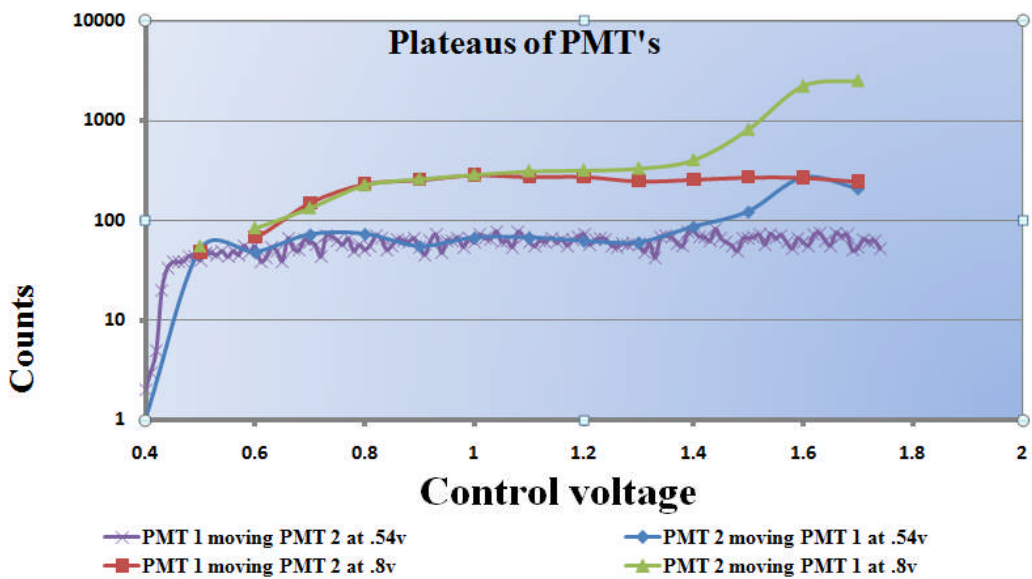

Figure 2. Coincidence counts per minute versus voltage. By fixing gain voltage for one and varying the voltage of the other PMT, a plateau region is identified. The range from 0.7 volt to 1.4 volt is found to be the best operating voltage for the two PMTs. Throughout the paper, all measurements and data reported were for 1 volt setting.

When observing the coincidental count rates, we set one PMTs at a fixed control voltage and adjusted the rate of the other PMT. As illustrated in Fig. 2, when the control voltage of one PMT was set at 0.54 V and the other PMT was adjusted through the range of control voltage settings, a long plateau region was observed.

Once the plateau regions were identified, the detector modules were positioned at different spacings and we recorded the number of coincidence counts that the radioactive sources produced. We also used different types of shielding, such as lead, aluminum and wood, to observe how well each piece of shielding worked, further examining the overall consistency and functionality of the detector. The detector modules were then set at 90-degree angles in relation to each other and recorded the number coincidence counts in the presence of a radioactive source as in Fig. 3.

To better understand muons-flux direction, we measured the number of coincidence counts versus detector orientation. The initial configuration was such that the detector pointed vertically; the angle was then varied. One set of measurements were taken in the laboratory and another set was taken in the field.

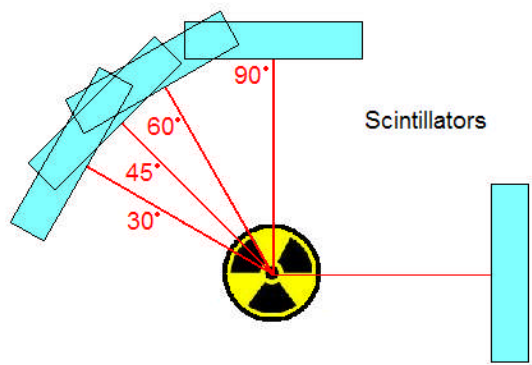

Figure 3. Radioactive source versus detector setup. The radioactive source and one of the modules is kept fixed while the other is rotated.

**Results**

*i. Cobalt 60*

When testing the cobalt-60 source, both modules were aligned sideways and their distance from the source varied (Fig. 4). The number of counts was plotted as a function of distance from the source. The results are shown in Fig. 5.

An angular measurement was also taken with one of the PMT tilted on its side and the other PMT

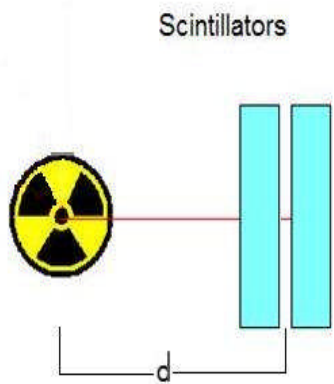

Figure 4. Experimental setup examining the inverse-square-law of radiation. The two scintillators were kept at a fixed position, while the distance *d* from the source varied.

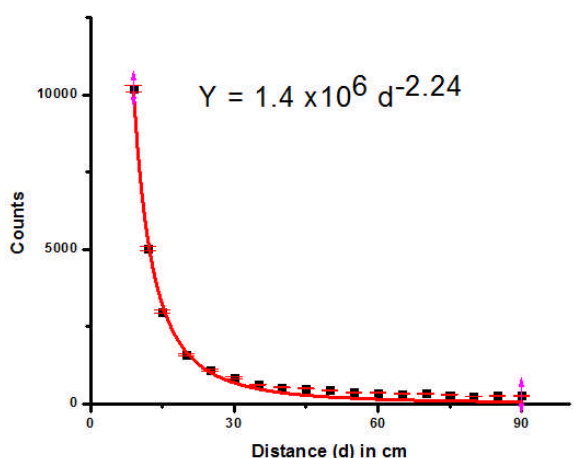

Figure 5. Y=1.39606x$10^6$ $d^{-2.24}$ This is the function for the graph displaying Counts vs. Distance (d) of the radiation-attenuation law. The graph refers to gamma rays produced by cobalt-60.

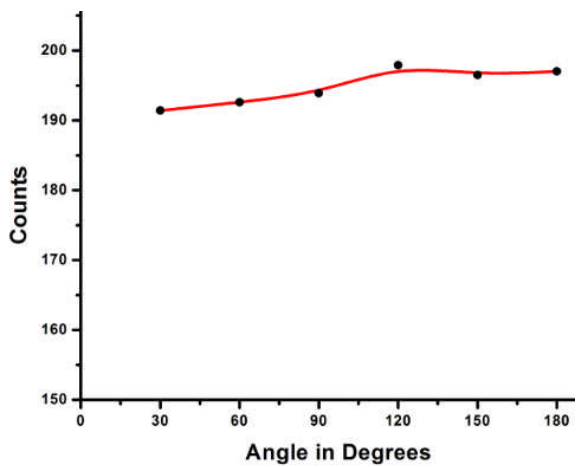

Figure 6. Graph illustrating the uniformity of counts. Result show that gamma rays from $^{60}$Co are emitted isotropically. Source-detector configuration used corresponds to that of Fig. 3.

revolved around different angles. Coincidence counts were taken in the presence of a $^{60}$Co source.

In testing different types of shielding, a measurement with no shield was taken 25.4 cm from the cobalt source and the PMT; and a record of 37051 counts per minute was measured. Next a 0.3175-cm aluminum shield was placed between the source and PMT and a count of 35828 was registered. Next, the aluminum shield was replaced with a 5.08-cm wood block and a count of 35630 was detected. After that a single and then another 6.35-cm lead acid battery was added between the source and the detector and counts of 22826 and 16828 were recorded.

### ii. Cosmic-Ray Muons

The PMTs are very sensitive and generate their own noise. This is largely due to thermal energy causing an electron to be ejected, which then excites several other electrons in the neighboring plates, triggering a false signal. To minimize the noise, a plateauing procedure was performed as described earlier for each PMT module. The procedure was done while looking for coincidence events, which revealed a plateau range around one volt without major disruption in the number of counts between different control voltages.

After the PMTs were plateaued, they were set to a vertical configuration and then tilted to determine the count rates at various degrees from horizontal (Fig. 7).

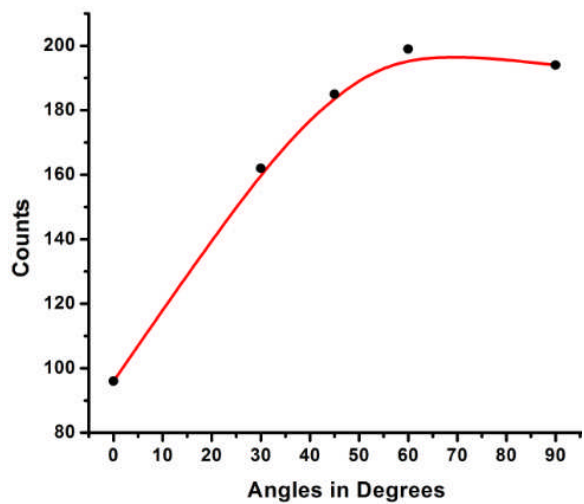

Figure 7. Count of CR flux vs. the angle from horizontal. A 90° angle corresponds to the detector sitting straight up, while 0° is when it points to the horizon (see Fig. 1).

**Discussion and Conclusion**

The $^{60}$Co source produced results similar to the expected $\frac{1}{r^2}$ but not exactly. The experimental results were not as exact because the inverse square law is derived for an ideal point source subtended by a spherical-detection aperture, whereas the detector at hand uses rectangle-shaped scintillators. Due to its rectangular geometry, we expect the scintillator to lose some of the gamma rays (Taylor et al. 2010) that would have otherwise contributed to the inverse square law, explaining the apparent additional 0.24 attenuation in our findings. This is particularly apparent as our observation matches closely the $\frac{1}{r^2}$ dependence when the source gets closer to the PMT module. The angular investigation, however, agrees with the common knowledge that the gamma rays emitted from the $^{60}$Co source are isotropic as they leave the source. Our findings are such that the detection rate was the same when measurements were taken along axes in any direction. The CR experiments verify previous experiments demonstrating that as the PMT modules are moved closer together, the solid angle gets larger and the number of detected CRs increases. Detailed analysis of solid-angle dependence will be reported elsewhere. The CR angular measurements suggest that, since the muon flux decreases as the two stacked modules are moved away from a vertical arrangement, the muons mostly come from above and shower straight down to Earth, rather than showering at some inclination angle. Another possible reason for this recorded attenuation in flux when the PMT's axis is closer to horizontal, is that more of the muons would have decayed because of longer trips in the atmosphere to get to the detector when it points sideways.

The end of this project is just the beginning for these detectors. Through the experimental process, the equipment underwent many repair procedures including the re-cementing of the light guides to the photomultipliers. Several of these came loose through the use of the modules. The mechanical device needs to be improved to hold the scintillator and light guide to the PMT. The mechanism built to separate the two modules from each detector is difficult to adjust and bends, not allowing the modules to be held together in perfect alignment.

Once these issues are resolved, the detector can be used to train future physics students in the detection of CR muons and gamma rays. Investigating the radiation-flux dependence on altitude in the state of Arkansas will greatly contribute to the scientific knowledge about the state; this will be explored in future work. The greater collection of data that will be accumulated by students over the next several years with these detectors will give other students and faculty of other disciplines a chance to understand the processes that are involved all around them. We hope that these detectors will not only serve physics undergraduates, but rather be used as a tool to incite the imagination of high-school students who are interested in physics.

**Acknowledgments**

We would like to thank everyone in association with the building of the detector, Dr. Helmuth Spieler, Dr. Howard Matis and Dr. James Siegrist. Special thanks to LBNL, Center for Science and Engineering Education of LBNL, Physics Division of LBNL. This work has been supported by the Department of Energy and National Science Foundation.